\documentclass[preprint]{aastex}
\input epsf

\def\cc{{\rm cm ^{-3}}}
\def\mum{\mu {\rm m}}

\def\lsun{{\,L_\odot}}

\def\ffir{{\rm F_{\rm FIR}}}
\def\lfir{{\rm L_{\rm FIR}}}
\def\rat{{\rm L_{\rm [CII]}}/{\rm L_{\rm FIR}}}
\def\2rat{{\rm L_{\rm [CII]}}/{\rm L_{\rm CO}(1-0)}}

\def\lb{{\rm L_{\rm B}}}
\def\6f{{\rm F_\nu(60 \mum)}}
\def\f100{{\rm F_\nu(100 \mum)}}
\def\msun{M_{\odot}}
\def\lsun{L_{\odot}}
\def\HI{\ion{H}{1}}
\def\CII{\ion{C}{2}}
\def\OI{\ion{O}{1}}
\def\tnm{\tablenotemark}

\begin{document}

\title{Probing the Interstellar Medium in Early type galaxies with ISO observations}
\author{  S. Malhotra\altaffilmark{1,2,3}
D. Hollenbach\altaffilmark{4} 
G. Helou\altaffilmark{5} 
N. Silbermann\altaffilmark{5}  
E. Valjavec\altaffilmark{5} 
R.H. Rubin \altaffilmark{4} 
D. Dale\altaffilmark{5} 
D. Hunter\altaffilmark{6} 
N. Lu\altaffilmark{5} 
S. Lord\altaffilmark{5} 
H. Dinerstein\altaffilmark{7} 
H. Thronson\altaffilmark{8}
}

\begin{abstract}

Four IRAS-detected early type galaxies were observed with the Infrared
Space Observatory. With the exception of the 15 $\mum$ image of
NGC~1052, the mid-IR emission from NGC~1052, NGC~1155, NGC~5866 and
NGC~6958 at 4.5, 7 and 15 $\mum$ show extended emission. Mid-IR
emission from NGC~1052, NGC~1155, and NGC~6958 follows a
de~Vaucouleurs profile. The ratio of 15/7 $\mum$ flux decreases with
radius in these galaxies, approaching the values empirically observed
for purely stellar systems. In NGC~5866, the 7 and 15 $\mum$ emission
is concentrated in the edge-on dust lane.

All the galaxies are detected in the [CII](158 $\mum$) line, and the
S0s NGC~1155 and NGC~5866 are detected in the [OI] (63 $\mum$) line as
well. Previous detections of neutral ISM are sparse: only NGC~1052 had
been detected in \HI\ and NGC~5866 in CO.  The ISO-LWS observations of
the [CII] line are more sensitive measures of cool, neutral ISM than
\HI\ and CO by about a factor of 10-100.  Comparison of [CII] with
H$\alpha$ shows that [CII] does not arise from HII regions and
therefore must arise in the neutral medium. Three of four early type
galaxies, namely NGC~1052, NGC~6958 and NGC~5866, have low ratios of
far-infrared to blue luminosity and show a lower $\rat$, which is
explained by postulating a softer radiation field from old stellar
populations in early type galaxies, compared to spirals and
irregulars, where young stars are present. While optical photons are
effective in heating the dust, UV radiation is needed to heat the gas
by the grain photoelectric mechanism. The low [CII]/CO ratio in
NGC~5866 ($\2rat \le 570$) confirms this scenario.

We estimate the UV radiation expected from the old stellar populations
in these galaxies and compare it to that needed to heat the gas to
account for the cooling observed [CII] and [OI] lines. In three out
of four galaxies, NGC~1052, NGC~5866 and NGC~6958, the predicted UV
radiation falls short by a factor of 2-3 of that required to
sufficiently heat the gas. In view of the observed intrinsic scatter
in the ``UV-upturn" in elliptical galaxies and its great sensitivity
to age and metallicity effects, this difference is  not
significant. However, the much larger difference (about a factor of 20)
between the UV radiation from old stars and that needed to produce the
far-infrared lines for NGC 1155 is strong evidence for the presence of
an additional UV source, probably young stars, in NGC~1155.

\keywords{radiation mechanisms: thermal, ISM: atoms, galaxies: ISM, infrared: ISM: lines and bands, infrared: ISM: continuum, galaxies: elliptical and lenticular, galaxies: ISM}
\end{abstract}

\altaffiltext{1}{Kitt Peak National Observatory, P.O. Box 26732, Tucson, AZ85705}
\altaffiltext{2}{Hubble Fellow}
\altaffiltext{3}{Now at Johns Hopkins University, Baltimore, MD 21218}
\altaffiltext{4}{NASA/Ames Research Center, MS 245-3, Moffett Field, CA 94035}
\altaffiltext{5}{IPAC, 100-22, California Institute of Technology, Pasadena,
CA 91125} 
\altaffiltext{6}{Lowell Observatory, 1400 Mars Hill Rd., Flagstaff, AZ 86001}
\altaffiltext{7}{University of Texas, Astronomy Department, RLM 15.308, Texas, Austin, TX 78712}
\altaffiltext{8}{NASA Headquarters}

\section{Introduction}

A significant fraction of early type galaxies (ellipticals and S0s)
contain some cool interstellar medium (ISM) including neutral gas and
dust (Jura et al. 1987, Knapp et al. 1989). However, the physical
conditions in the ISM in these galaxies are relatively unknown.  Many
ellipticals have been detected by the X-ray emission of hot gas
(Roberts et al. 1991), and many in the \HI\ and CO lines indicating
cold-neutral medium (Lees et al. 1991, Wiklind, Combes \& Henkle
1995), and through dust obscuration (Sadler \& Gerhard 1985,
Goudfrooij et al. 1994a 1994b). The environment in the ellipticals may
lead to different equilibrium physical conditions in the various
phases of the ISM because of the presence of hot X-ray gas and a
higher ratio of stellar radiation compared to the cool gas and
dust. But the IRAS colors in 60 and 100 micron emission indicate cold
dust with temperatures comparable to found in spirals.

We observed four elliptical/S0 galaxies with the Infrared Space
Observatory (Kessler et al. 1996) to study the physical conditions in
the gas and dust in these galaxies. These galaxies are part of a large
sample of normal star-forming galaxies spanning a range of
morphologies, infrared luminosities, FIR colors
($F_\nu(60)/F_\nu(100)$) and FIR-to-Blue ratio.  ISO-CAM (Cesarsky et
al. 1996) images at 7 and 15 microns are used to determine the
distribution of mid-IR emission (section 3) and hence its origin.  The
mid-IR spectrum of NGC~5866 was taken with ISO-PHT (Lemke et al. 2000)
and is presented by Lu et al. (2000).  The FIR spectroscopy of fine
structure lines: [CII] (158 $\mum$) and [OI] (63 $\mum$), was carried
out with ISO-LWS (Clegg et al. 1996).  Assuming [CII] and [OI] to be
the major cooling lines, we estimate gas heating rates and examine
possible heat sources in these galaxies in section 4. For NGC~1155 and
NGC~5866 where both [CII] (158 $\mum$) and [OI] (63 $\mum$) lines were
detected we estimate the radiation and gas density, $G_0$ and $n$.

The sample considered in this paper consists of two elliptical
galaxies: NGC~6958 and NGC~1052 and two S0 galaxies: NGC~1155 and
NGC~5866. Like any sample of four galaxies, these have varied
properties. NGC~1052 harbors an active nucleus (Fosbury et
al. 1978). NGC~5866 has a large edge-on disk.  NGC~1155 is labelled an
elliptical in the NASA Extragalactic Database (NED), but has been
reclassified as an SO by Harold Corwin (private communication) based
on morphology using B-band CCD images taken at Lowell
Observatory. These images were taken with a CCD camera with
0.61$\arcsec$ pixels and approximately 1.5$\arcsec$ seeing. This
galaxy is small (0.8 $\arcmin$ ) and distant which may make a
definitive classification difficult. The B-band surface brightness 
profile is well fit by a de~Vaucouleurs profile and cannot be
fit with a point source plus exponential disk. NGC~1155 is also
labelled a starburst in NED and is a Markarian galaxy. It has a spiral
companion (NGC~1154) two arcminutes away and a bridge can be seen
between the two galaxies in deep optical images.  NGC~6958 seems the
most unremarkable galaxy in this set.

\section{Observations and data analysis}


The Mid-IR imaging of NGC~1155 and NGC~6958 was done with ISO-CAM at
$6.75 \mum$ (LW2 filter, $\Delta \lambda =3.5 \mum$) and at 15 $\mum$
(LW3 filter, $\Delta \lambda = 6 \mum$), using the raster scan mode to
cover roughly $ 4.5 \arcmin \times 4.5 \arcmin$ centered on the
nucleus.  CAM was set to $6 \arcsec$/pixel, and the raster was made up
of $2 \times 2$ pointings separated by $81 \arcsec$, or 13.5 pixels in
each direction, allowing for better spatial sampling.  At each
pointing in the raster scan we integrated for 50 seconds in each band:
$25 \times 2$ second exposures at 15 $\mum$, and $10 \times 5$ seconds
at 7 $\mum$. Typical sensitivity is 0.02 Jy/pixel in the $7 \mum$
filter and 0.04 Jy/pixel in the $15 \mum$ filter.  More details of CAM
data reduction can be found in Dale et al (2000). NGC~1052 and
NGC~5866 were observed by Vigroux et al. in the LW1 (4.5 $\mum$), LW2
($7 \mum$) and LW3 ($ 15\mum$) filters with 3 and 6 $\arcsec$/pixel. A
detailed analysis of these data is presented by Madden et al. 2000.


With the $80 \arcsec$ beam of the long wavelength spectrograph (LWS)
and spectral resolution of $0.6-0.3 \mum$ we measure total line flux
for these galaxies.  The galaxies are estimated to have $FWHM <
0.5\arcmin$ in FIR emission using deconvolved IRAS maps. The [CII]
line observations were planned to achieve ($1\sigma$) sensitivities of
$5\times 10^{-5}\times \ffir$, where $\ffir$ is the total far-infrared
flux of the galaxy and is computed according to the relation $
\ffir=1.26 \times 10^{-14} [2.58 \times F_{\nu}(60\mum) +
F_{\nu}(100\mum)] W m^{-2}$ (Helou et al. 1988), where
$F_{\nu}(60\mum)$ and $F_{\nu}(100\mum)$ are flux densities in Jansky
at 60 and 100 $\mum$ measured by IRAS.

The data were reduced and calibrated with the ISO data reduction
pipeline OLP7.0.  Post-pipeline data reduction and recalibration was
carried out using an interactive data reduction package ISAP
(cf. http://www.ipac.caltech.edu/iso/lws/lws.html).  The line profiles
were derived from several scans by running a median boxcar filter
through the scans.  We use the median of the observed fluxes instead
of the mean to reduce the influence of outlying points arising from
cosmic ray hits.  Line fluxes were derived by integrating directly
under the lines, after fitting a linear baseline to the continuum (cf
Malhotra et al. 2000 for details on the LWS data reduction and line
fluxes).

Optical imaging of NGC~1155 was done at Lowell Observatory.  The
H$\alpha$ images were obtained through narrow-band filters centered at
the line; an off-band filter was used to image and subtract the
continuum. Long slit spectroscopy and H$\alpha$ and broad-band imaging
of NGC~1155 was done at Lowell observatory. CO(1-0) observations on
NGC~6958 were carried out at the Swedish ESO Submillimeter Telescope
(SEST).

\section{Mid-infrared imaging} 

In spiral galaxies the mid-IR emission is dominated by small grains
transiently heated to high temperatures and the fluorescence of large
aromatic molecules (Puget \& Leger 1989).  This may not be the case
for ellipticals given that the ISM-to-stellar ratio is low. A
significant part of the emission at 12 microns could be from
photospheres and circumstellar dust (Knapp, Gunn and Wynn-Williams
1992), although Sauvage and Thuan (1994) have argued that the IRAS
color-color relation (Helou 1986) holds for early type as well as for
late type galaxies indicating a similar origin from ISM. Hot dust near
an active nucleus (e.g. AGN) can also be a major contributer in some
cases. Because of its high spatial resolution and sensitivity, CAM
imaging at 7 and 15 $\mum$ can tell us about the spatial distribution
of the mid-IR emission in galaxies.  The interpretation is complicated
by the fact that the mid-IR emission maps depend on the distribution
of the dust as well as the heating sources.

\begin{figure}[htb]
\centerline{\epsfxsize=3.0in \epsfbox{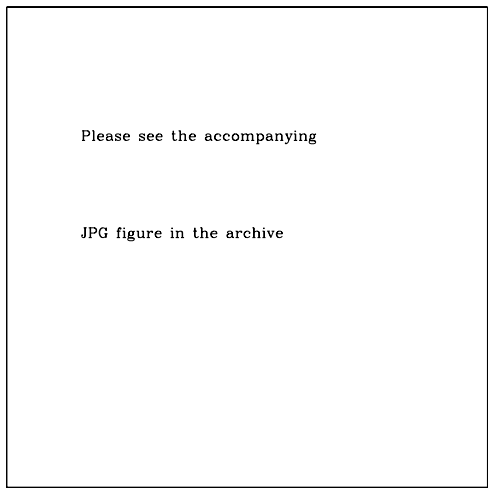} }
\caption{ The mid-infrared and optical images of NGC~1052, NGC~1155,
NGC~5866 and NGC~6958 are presented for comparison. The images are
(going left to right): DSS (digitised sky survey) images in R, 4.5
$\mum$, 6.75$\mum$ and 15$\mum$ images from CAM. The 4.5$\mum$ images
of NGC~1052 and NGC~5866 trace the starlight, while the 7$\mum$ and
15$\mum$ images of NGC~5866 trace the edge on dust lane. The 7$\mum$
and 15$\mum$ images of NGC~1052 are dominated by the active
nucleus. All except the 15 $\mum$ image of NGC~1052 show extended
emission. The orientation of the images is
indicated by the arrows on the DSS images, where the arrowhead points
north. The field of view is $4.5\arcmin$ on the side for NGC~1155 and
NGC~6958 and $1.5 \arcmin$ on the side for NGC~1052 and NGC~5866 }
\end{figure}

The mid-IR emission is extended in at least 9 out of 10 mid-IR images
(Figure 1), although many of the images look compact because the light
in ellipticals is very concentrated. Comparison of curves of growth
(i.e. flux enclosed in a radius) show significant difference between a
point source and these galaxies (Figure 2) except for the 15 $\mum$
image of NGC~1052. 

The 4.5 $\mum$ emission from NGC~1052 is extended and traces the
stellar distribution. At 7 and 15 $\mum$ NGC~1052 comes closest to
being dominated by the central point source, which is an active
nucleus in this case.  The flux in 7 and 15 micron bands in the
central 6 $\arcsec$ radius in NGC~1052 is about 80\%, compared to 88\%
and 80 \% expected for a point source in 7 and 15 micron filters with
6 $\arcsec$ pixel-scale. There is some uncertainty in the expected PSF
depending on the placement of the point source within a 6$\arcsec$
pixel. NGC~1052 was also observed in the 3 $\arcsec$/pixel mode on CAM
which samples the PSF properly. The images with 3 $\arcsec$/pixel show
that at $7 \mum$ NGC~1052 is extended and at 15 microns it is
consistent with being unresolved (Figure 2). In this respect it is
similar to NGC~3998 whose mid-infrared emission is dominated by the
active nucleus (Knapp et al. 1996). The 7 and 15 $\mum$ emission is
extended for NGC~1155, where the central 6$\arcsec$ accounts for only
47\% and 56\% of the total flux at 7 and 15 $\mum$. In NGC~5866 the
central 6$\arcsec$ contains 24\% and 40\% of the total flux in 7 and
15 $\mum$ bands.  In NGC~5866 the mid-IR emission at 7 and 15 $\mum$
traces the nearly edge-on dust lane, whereas the 4.5$\mum$ emission
traces the stars (Figure 1).

\begin{figure}[htb]
\centerline{\epsfxsize=5.0in \epsfbox{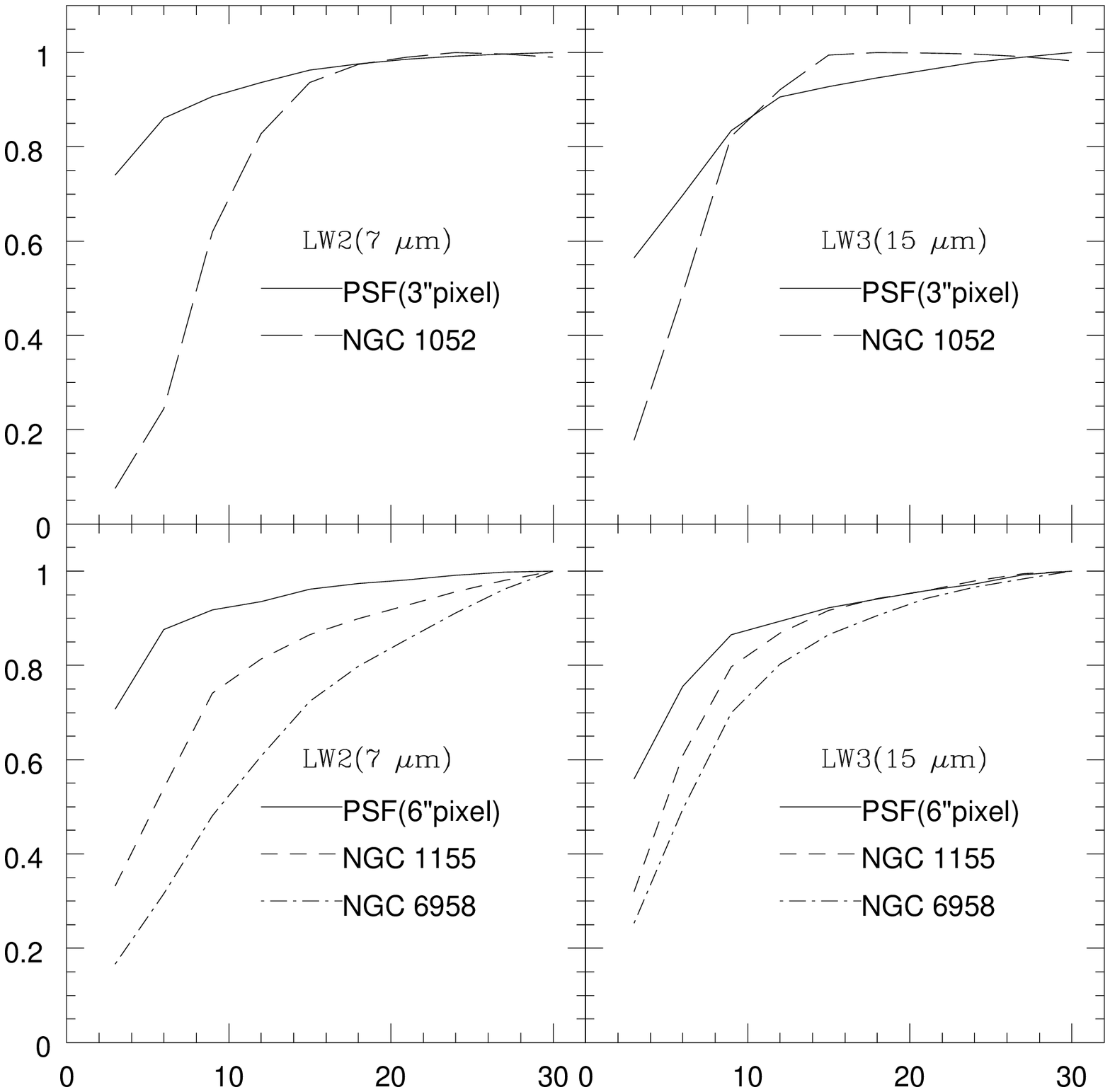} }
\caption{ This plot shows the cumulative flux profiles of elliptical
galaxies at 7 and $15 \mum$ as compared to the point source
profiles. NGC~1155 and NGC~6958 were observed in a mode with 1 pixel=6
$\arcsec$ and they are plotted in the lower two panels. NGC~1052 was
also observed in the 3 $\arcsec$ pixel mode and its profile is plotted
in the upper two panels. Except for the 15 $\mu m$ image of NGC~1052,
all other images are clearly extended. This is manifest from the slow
rise of cumulative flux with radius for galaxies as compared to the
point source profile. Within measurement errors the 15 $\mu m$ profile
of NGC~1052 is consistent with an unresolved source.}

\end{figure}

\begin{figure}[htb]
\centerline{\epsfxsize=5.0in \epsfbox{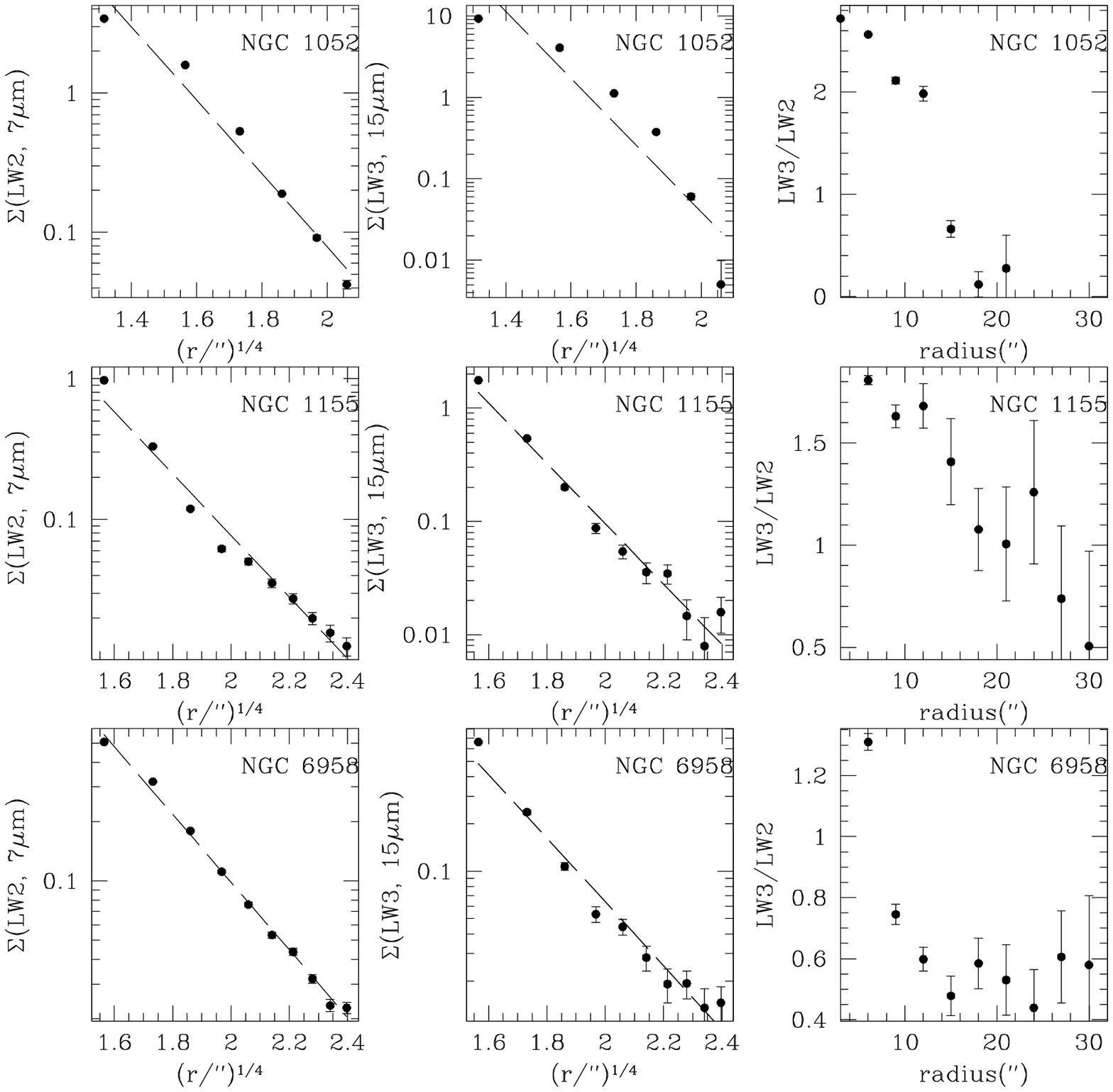} }
\caption{ The surface brightness of early type galaxies in the
mid-infrared at 7 and 15 microns follows de~Vaucouleurs radial
profiles (left and middle columns). The ratio of LW3 (15 $\mum$) and
LW2 (7 $\mum$) light decreases towards the outer parts of galaxies
(right column). The $15 \mum$/$7 \mum$ ratio for NGC~6958 reaches an
asymptotic value of 0.5, characteristic of old stellar populations in
galaxies devoid of ISM. This indicates that the ISM is concentrated
near the center.}
\end{figure}

To increase the signal-to-noise in the outer parts of the galaxies we
average the emission in annuli after subtracting the background
determined at radii greater than $90\arcsec$. Figure 3 shows the
radial profile of the surface brightness in three galaxies, NGC~1052,
NGC~1155 and NGC~6958. The radial surface density profile is derived
by averaging azimuthally. The error-bars in surface brightness in
Figure 3 are derived from the scatter in pixel brightness in a given
annulus.  A de~Vaucouleurs radial profile provides a reasonable fit to
the mid-IR surface brightness profile.  The effective radius for the
mid-IR emission is intermediate to that seen in optical B-band and
H$\alpha$ light for NGC~1155 (Figure 4). We also note that in all
three galaxies the $15 \mum$/$7 \mum$ ratio decreases going from the
center of the galaxy outwards (Figure 3).

The $15 \mum$/$7 \mum$ ratio can be used as a further diagnostic to
determine the origin of the mid-infrared emission in the current
sample. In quiescent spiral galaxies this ratio is 1 (e.g. NGC~6946,
Helou et al. 1996). In more actively star-forming regions/galaxies the
$15 \mum$/$7 \mum$ ratio increases due to higher heating radiation
(Vigoroux et al. 1999, Dale et al. 1999). $15 \mum/7 \mum < 1$ is
found in early type galaxies where the $7 \mum$ band contains a
significant contribution from stellar photospheres (Madden et
al. 1997).  In the inner parts of NGC~1052 and NGC~1155, $15 \mum$/$7
\mum$ is a factor $\simeq2$ higher than the ISM in quiescent spiral
galaxies (Figure 3).  This is presumably due to higher radiation
density in the central regions of these galaxies due to higher density
of stars or the active nucleus in the case of NGC~1052.  The decrease
in $15 \mum$/$7 \mum$ in the outer parts of the galaxy is due to
different proportions of photospheric and ISM contributions, with the
photospheres having a lower $15 \mum$/$7 \mum$ ratio (i.e., bluer
emission).  This scenario is consistent with the $15 \mum$/$7 \mum$
ratio of 0.5 seen in the outer parts of NGC 6958. This ratio is within
the range of observed ratios in ellipticals devoid of ISM (Madden et
al 1997), and indeed NGC 6958 does not show extended H$\alpha$
emission beyond its nucleus, indicating that the ISM is not very
extended in this galaxy (Phillips et al. 1986) .

\begin{figure}[htb]
\centerline{\epsfxsize=3.3in \epsfbox{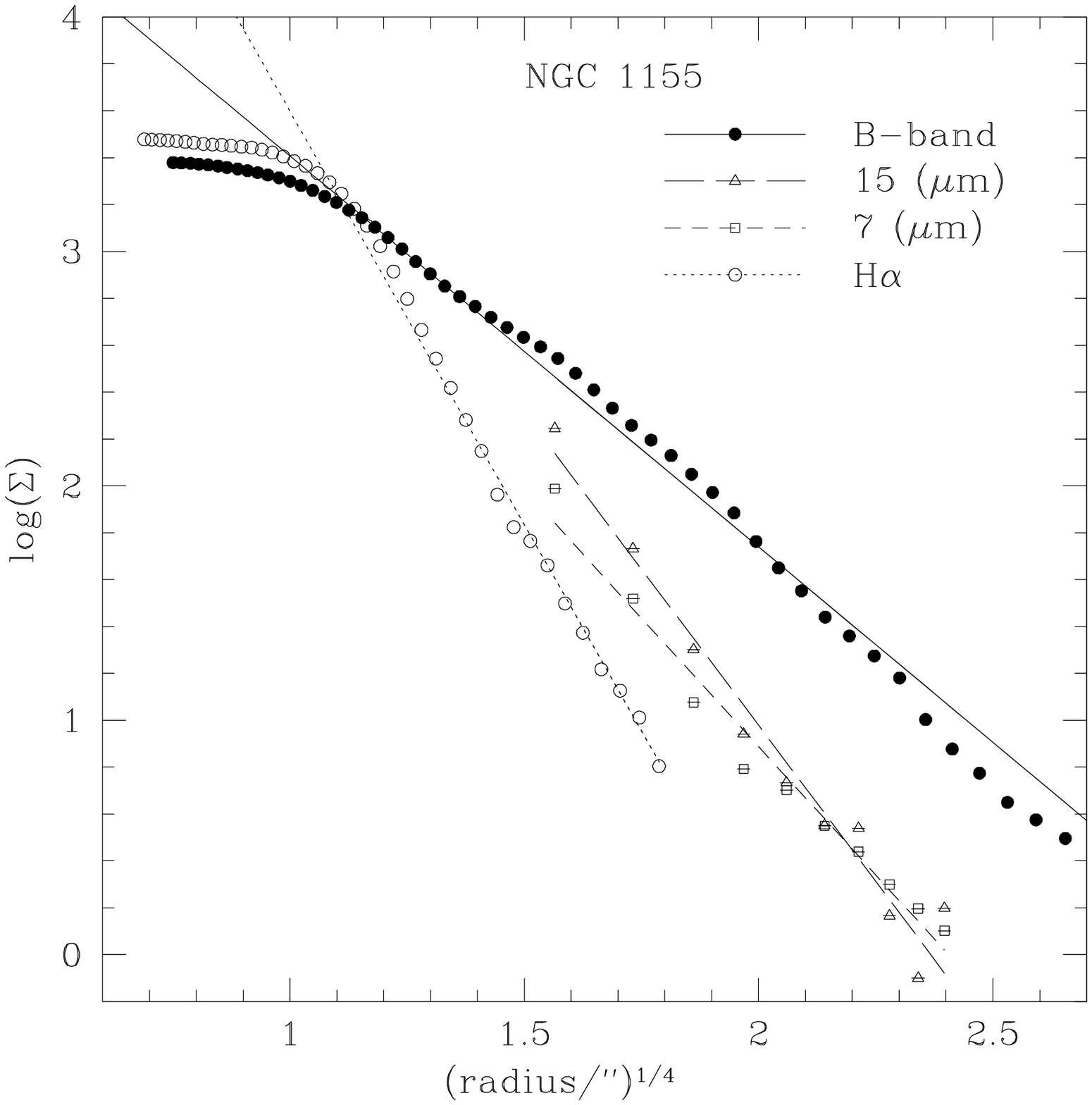} }
\caption{ The radial profile of optical B band emission (filled circles),
the H$\alpha$ emission (open circles), and the mid-infrared emission
($7 \mum$: open squares, $15 \mum$: triangles) in NGC~1155 is plotted along with
the best fit de~Vaucouleurs profile for each wavelength. The
mid-infrared light shows a radial distribution intermediate to the
B band and the H$\alpha$.}
\end{figure}

The spatial distribution of the mid-IR emission could also reflect the
distribution of sources heating the dust. In case of NGC~1155 the
mid-IR emission follows a de~Vaucouleurs profile with an effective
radius $R_e$ which is intermediate between H$\alpha$ and B band
indicating that both ionizing and non-ionizing radiation heat the
dust.  A similar result was found for the spiral galaxy NGC~6946
(Malhotra et al. 1996). If so, we would predict that the dust is quite
extensive in NGC~1155.  While the extent of the ISM is not known for
NGC~1155, the luminosity in the [CII] line is as much as a third of
the Milky Way value indicating the presence of a significant amount of
gas. The presence of an extended component of dust can explain the
discrepancy between dust masses obtained from optical extinction and
FIR emission, the former being sensitive only to dust concentrated
near the center (Goudfrooij et al 1994).

\section{Far-Infrared Spectroscopy} 

The physical conditions in the different phases of the ISM in early
type galaxies are not very well known, partly because there is not as
much cool ISM in these galaxies, and partly because we do not live in
such a galaxy. Much of our detailed knowledge about the various phases
of the ISM in spirals comes from being in these phases and able to do
absorption studies. In ellipticals/S0s hot gas is detected in the X-ray,
and ionized gas is detected by H$\alpha$ emission. The cold neutral
medium is detectable as \HI\ (21 cm) and CO by present instrumentation
for typical masses of $10^7-10^8 \msun$.  As a result the detection
rate for cold gas is about 15-50\% (van Gorkom 1997, Rupen 1997)
depending on the morphological classification of galaxies and
preselection of samples.  The detection rate is higher for H$\alpha$
(about 50\%) where one is sensitive to $10^5 \msun$ of gas (Phillips
et al. 1986).

The far infrared fine structure lines dominate the global cooling of
the neutral ISM of galaxies.  Of these, [CII] (158 $\mum$) and [OI] (63
$\mum$) are generally the strongest, with [CII] dominating the lower
densities and moderate UV radiation field intensities (Hollenbach,
Tielens \& Takahashi 1991; Hollenbach \& Tielens 1997, 1999). We
detect the [CII] transition in all four galaxies, albeit at low
(3-$\sigma$) S/N for NGC~6958. The other major fine structure cooling
line, [OI] (63 $\mum$), is detected in NGC~1155 and NGC~5866. In Table
1 we summarize what is known about the warm and cool ISM in the four
early type galaxies. Detections are few - \HI\ (21 cm) is detected in
NGC~1052 (van Gorkom et al. 1986), CO in NGC~5866 (Thronson et
al. 1989) and H$\alpha$ is detected in all four. There is tentative
detection of CO in absorption against the nucleus in NGC~1052 (Knapp
\& Rupen 1996).  In Table 1 $M_H(C^+)$ is the minimum mass
of the hydrogen associated with the observed C$+$ emission and is
derived in the high density, high temperature limit ($n_H> 3000, T \gg
91K$), assuming solar abundance of carbon with all the carbon in the
form of C$+$ (Crawford et al. 1985). $M_H(C^+)$ is typically 10-100 times
smaller than the observed detection or upper limits on HI and H$_2$.

\begin{table}[htb]{}
\caption{Mass in Interstellar components}
\begin{tabular}{lccccccl}
\hline\noalign{\smallskip}
Name & Morphology \tnm{j} & Distance & M(HI) & $M(H_2)$ & M(HII)& $M_H(C^+)$&$M_{dust}$\tnm{i}\cr
&&& (Mass& in & $\msun$) &  \cr
\noalign{\smallskip}
\hline\noalign{\smallskip}
NGC~1052 & E4    & 14 Mpc   &$ 0.7 \times 10^7$ \tnm{a} &$<3\times10^7$ \tnm{b}&$3\times10^4$ \tnm{c} &$>4 \times10^5 $&$4.8\times10^4$\cr
NGC~1155 & S0 & 45 Mpc   &    ?             	       &     ?                 &     ?                &$>1 \times10^7 $&$1.8\times10^6$\cr
NGC~5866 & S0    & 11.4 Mpc &$ < 3 \times 10^7$ \tnm{f} &$>3\times10^8$ \tnm{g}&$ 4\times10^4$ \tnm{h}&$>9.7\times10^5$&$1.2\times10^6$\cr
NGC~6958 & E+    & 28 Mpc   &$  <3 \times 10^8$ \tnm{d} &$<9\times10^8$ \tnm{e}&	?  	      &$>8 \times10^5 $&$3.5\times10^5$\cr
\noalign{\smallskip}
\hline
\end{tabular}
\tablenotetext{a}{from van Gorkom et al. 1986, adjusted to the LWS beam-size of $70 \arcsec$}
\tablenotetext{b}{Knapp \& Rupen 1996, Wiklind et al. 1995}
\tablenotetext{c}{Plana \& Boulesteix 1996}
\tablenotetext{d}{Walsh et al. 1990}
\tablenotetext{e}{recent observations at SEST}
\tablenotetext{f}{Haynes et al. 1990}
\tablenotetext{g}{Thronson et al. 1989}
\tablenotetext{h}{Plana et al. 1998}
\tablenotetext{i}{Roberts et al. 1991, adjusted for the given distances}
\tablenotetext{j}{from NASA Extragalactic Database and Harold Corwin(private communication)}
\end{table}

\subsection{ Trends in line strengths in Early type galaxies}

The [CII] line strengths of normal spiral and irregular galaxies
follow well defined trends with far-infrared colors ($\6f/\f100$)
indicating dependence on dust temperature and therefore on dust
heating radiation density; and on star-formation activity, measured by
the ratio of far-infrared and blue band luminosity $\lfir/\lb$.  The
ratio $\rat$ decreases dramatically with increasing $\6f/\f100$ and
$\lfir/\lb$ (Malhotra et al. 1997, 2000).

While [CII] and [OI] are the main coolants of warm ($\simeq 90$ K),
neutral gas; the main source of heating are the photoejected electrons
from dust grains (Watson 1972). Only UV light is effective in
producing photoejected electrons since the potential barrier for
neutral grains is approximately 6 eV, or more if the grains are
positively charged (Bakes \& Tielens 1994).  The far-infrared
continuum reflects the cooling, and thus the total heating of
the dust grains. For galaxies with the most active star-formation and
warm dust and hence most intense radiation fields, the dust grains
become positively charged making the photoelectric heating less
efficient. Therefore we see a decrease in [\CII] and [\OI] lines
strengths relative to the dust continuum for the most active galaxies
with higher $\6f/\f100$ and $\lfir/\lb$ (Figure 5).


\begin{figure}[htb]
\plottwo{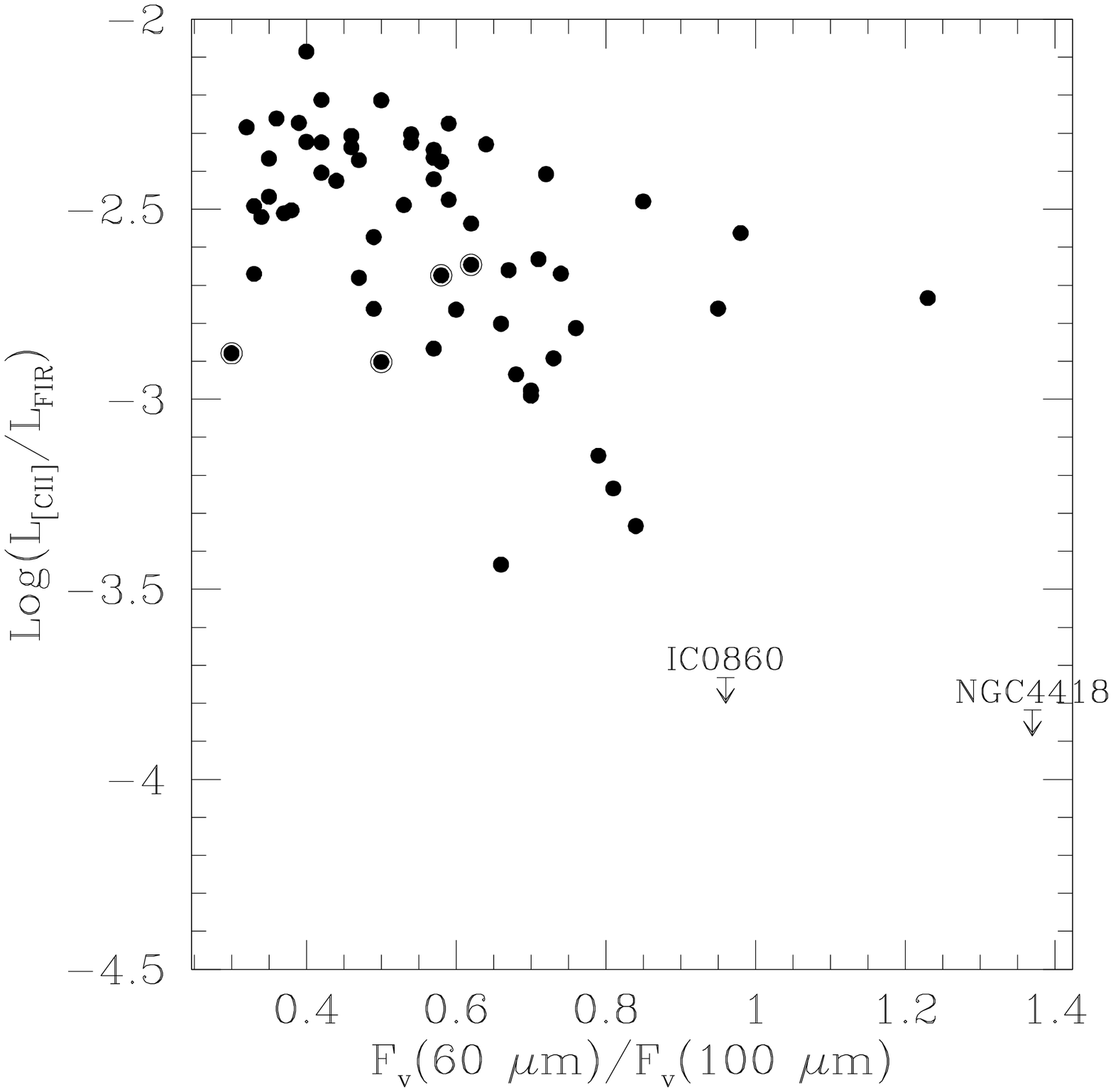}{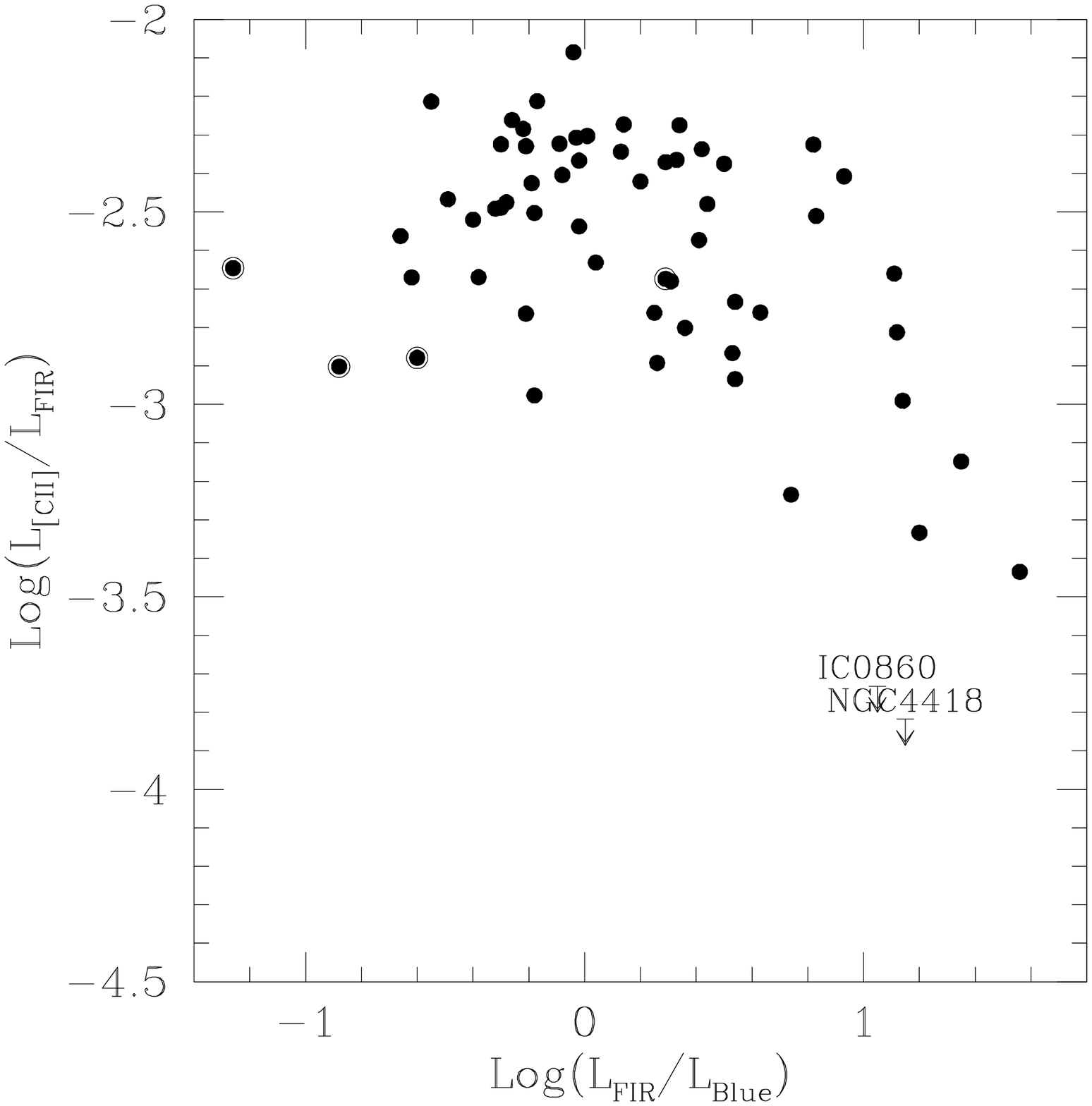}
\caption{ The ratio of FIR and [CII] line luminosities is plotted
against the dust temperature characterized by the ratio of flux
densities at $60 \mum$ and $100 \mum$ from IRAS measurements for a
sample of galaxies spanning a whole range of morphologies (cf Malhotra
et al. 2000). The four early type galaxies are shown with the bull's
eye symbol, and lie within the broad trend of decreasing $\rat$ with
the increase in $\6f/\f100$.  (b) $\rat$ shows a similar trend with
the ratio of FIR/Blue luminosity, which can be used as an indicator of
star formation activity. Three out of four early type galaxies are at
the lower end of mass normalised star-formation activity. These
galaxies-NGC~1052, NGC~6958 and NGC~5866 show a lower $\rat$ due to
softer radiation field from aging stellar populations.}
\end{figure}

On the other extreme, galaxies with low rates of current
star-formation, and softer radiation field a low $\rat$ is expected
since only the UV photons ($\lambda < 2000 \AA$) contribute to
photoelectric heating of gas, while dust is heated by optical as well
as UV photons. In early type galaxies, which typically have
lower active star-formation rate relative to the total blue
luminosity, we do see a lower $\rat$.

In Figures 5a and 5b we plot these trends for the whole sample of
distant galaxies with the early type galaxies circled. It is apparent
in these figures that the four Es and S0s lie within the broad
correlation between $\rat$ and $\6f/\f100$ and $\lfir/\lb$. NGC~6958
and NGC~1052 and NGC~5866 lie at the low end of $\lfir/\lb$,
indicating a low rate of current star-formation and low
extinction. The $\rat$ ratio turns down slightly at the low end of
star-formation.

\subsection{Origin of [CII]}
The [CII] emission could be from ionized gas in dense HII regions, diffuse
ionized gas, neutral atomic medium, or from UV irradiated surfaces of
molecular clouds. In nuclear regions of galaxies [CII] arises mostly
from dense PDRs (Stacey et al. 1991, Crawford et al. 1985). However,
integrated over the disks of normal spiral galaxies, a significant
fraction (up to 50\%) may also arise from diffuse medium, i.e. the
cold neutral medium (CNM) (Madden et al. 1993) or from extended, low
density ionized gas (Heiles 1994). With the limited knowledge of the
phases of the ISM in these galaxies and without assuming that the
temperature and density in these phases is very similar to the
Galactic values, it is hard to identify the phase of the ISM which
produces most of the [CII] emission.

We can rule out HII regions as sources of much
[CII].  Assuming case B recombination and $T_e=10^4$ K, one can place
an upper limit on [CII] emission from HII regions (Malhotra et al. 2000) as

$$\frac{\rm L_{\rm CII}}{\rm L_{\rm H\alpha}}< 5.4 \times 10^4 x_C
\bigg(\frac{1}{n_e+17}\bigg)$$ 
here $x_C$ is the gas phase abundance of carbon with respect to hydrogen, and $n
_e$ is the density of electrons in the HII region in $\cc$.
The inequality sign is due to the fact that not all of the C in HII regions 
exists as $C^+$; some will be in the form of $C^{2+}$.

For low electron density ($n_e \rightarrow 0$):
$$\frac{\rm L_{\rm CII}}{\rm L_{\rm H\alpha}}< 0.33 x_{-4}$$
where $x_{-4}=x_C/10^{-4}$
In the high density limit ($ n_e >> 17\,\cc$):
$$\frac{\rm L_{\rm CII}}{\rm L_{\rm H\alpha}}< 0.054 x_{-4} \bigg(\frac{10^2 \cc
}{n_e}\bigg)$$

For NGC~1155, NGC~5866 and NGC~6958 the ratio $L([CII])/L(H\alpha)$ is
measured to be 2.5, 0.5 and 58 respectively indicating that most of
the [CII] is not associated with HII regions. For NGC~1052 the observed
$L([CII])/L(H\alpha)=0.16$.  Taking $n_e = 10^3 cm^{-3}$, inferred
from the ratio of SII lines (Kim 1989), the contribution of HII
regions to CII emission is expected to be $L([CII])/L(H\alpha) <
0.015$ (for $x_{-4}=3$); an order of magnitude lower than the observed
levels. Thus it appears that not much of [CII] emission arises from
the ionized gas in early type galaxies. The total extinction is low so
the H$\alpha$ flux is unlikely to be extincted by more than a factor
of two.

Could the [CII] emission be coming from diffuse HI? It is hard to
answer this question given that HI column density has been measured
only in NGC~1052. For a cold neutral medium (CNM) with typical
Galactic values T=80K and $n_H=90 \cc$, $M(HI)/L(CII)= 20 (3 \times
10^{-4}/X_C{^+}) \msun / \lsun$. For NGC~1052 $M(HI)/L(CII)=14$ which
can easily be accommodated with a slightly higher gas density or
warmer gas, or some contribution to [CII] from molecular medium. It is
quite likely that most of the [CII] emission is associated with
molecular gas in NGC~5866, given that the upper limits on \HI\ mass
are ten times lower than the observed M(H$_2$). Unfortunately, no useful
constraints can be placed on the origin of [CII] in NGC~1155 and
NGC~6958 where neither \HI\ nor CO have been detected.  It should be
easy to estimate the \HI\ contribution to the origin of [CII] emission
in NGC~1155 with \HI\ observations. If [CII] is associated with \HI\
we should see $M(HI)\simeq 10^8 \msun$.

\subsection{[CII]/CO ratio}

For [CII] emission arising from photo-dissociation regions, softer
radiation leads to a lower fraction of gas in C$^+$ due to reduced
photodissociation of CO molecules, making a thinner C$^+$ layer on the
surface of the molecular cloud (Spaans et al. 1994). This scenario
predicts a lower ratio of the flux in [CII] versus various CO
transitions as well as lower $\rat$.  For Galactic HII and OB
star-forming regions and for starburst galaxies the ratio
$\2rat=4100$, whereas for more quiescent spiral galaxies this ratio
was found to be $\2rat=1500$ (Stacey et al 1991). For NGC~5866, which
is the only galaxy in the current sample with a CO measurement $\2rat
\le 570$.  This is an upper limit because the CO is measured in the
$45\arcsec$ beam of FCRAO (Thronson et al. 1989) and the [CII] is
measured in a $70 \arcsec$ beam. This provides evidence for a soft
radiation field. For NGC~6958 and NGC~1052 we only have upper limits
on CO(1-0) emission. For NGC~1052 $\2rat > 1100$, which is comparable
to the quiescent spiral galaxy sample, but most of the [CII] seems to
be associated with the atomic gas in this galaxy, while the low
$\2rat$ prediction in soft radiation fields is made for dense PDRs on
the surfaces on molecular clouds. In NGC~5866, where we find low
$\2rat$ values, the M(HI)/M(H$_2$)$<0.1$ indicating that most of the
[CII] is associated with molecular clouds, providing a cleaner
diagnostic of the softer radiation field. For NGC~6958 upper limits on
CO imply $\2rat > 170$ and more sensitive CO observations are needed.

\subsection{Physical conditions in the gas}

With the few FIR line measurements of these galaxies we can attempt
to derive the physical conditions in the warm, neutral, FIR emitting
gas. 

{\it NGC~1155:} We compare the two FIR line ([CII] and [OI]($63\mum$))
and line to continuum ratios in NGC~1155 to models of
photodissociation regions by Kaufman et al. 1999. Using the ratios
([CII]+[OI])/FIR and [OI]/[CII] gives density $n \simeq 10^2 \cc$ and
FUV flux $G_0 \simeq 10^2$ ($G_0$ conventionally is the FUV flux
normalized to the average local interstellar flux of $1.6 \times
10^{-3}$ ergs cm$^{-2}$ s$^{-1}$ (Habing 1968)) There are no HI or CO
measurements for this galaxy.  The chances of detecting both are
excellent, both because of the high minimum mass of hydrogen
associated with the $C^{+}$ emission and because of the density of the
gas derived.

{\it NGC~1052:} The column density of HI is measured to be $2.5 \times
10^{20} cm^{-2}$ (van Gorkom et al. 1986) near the nucleus and about 1
arcminute away.  It is then reasonable to assume that N(HI) is
constant in the LWS 70$\arcsec$ beam and the [CII] emitting gas fills
the beam: I([CII])$= 1.1 \times 10^{-6} ergs/cm^2/s/sr$. Using this in
the following equation with the further assumptions that the
temperature of the gas $T>91 K$ and the ionization fraction is small
($X_e < 10^{-3}$, so we can neglect collisional excitation due to
electrons), we can solve for the density of the gas $n_H$.

$$ I([CII])= 2.3 \times 10^{-21}\  N(HI) \ X_{C^+} \ ( \frac {2 e^{(-91/T)}} 
{1+2 e^{(-91/T)} + n_{crit}/n_H})$$

Where $n_{crit}=2\times 10^{3}$ is the critical density for
collisions with HI (Launay \& Roeff 1977, Hayes \& Nussbaumer 1984)
and $X_{C^+}$, the $C^+$ abundance is taken to be the Galactic value
of of $3\times 10^{-4}$. The density of the HI then is derived to be
$n_H=6 \cc$, consistent with [CII] arising from diffuse HI.

{\it NGC~5866} Comparing the models to the observed ([CII]+[OI])/FIR,
[CII]/[OI] and [CII]/CO(1-0) ratios does not yield a solution for gas
and radiation density in NGC~5866, because the models assume a harder
radiation field, and that half the dust heating is from far-UV. In
softer radiation fields from old stars the optical light contributes a
larger fraction of the dust heating and therefore to the FIR
continuum. 

Ignoring ([CII]+[OI])/FIR ratio, we estimate the intensity of [CII]
emission in the beam, assuming a beam filling factor of 1, to be
I([CII])$= 5.5 \times 10^{-6} ergs/cm^2/s/sr$. Using this and the
ratios [OI]/[CII]=0.2 and [CII]/CO(1-0)$ \le  570$, PDR models of
Kaufman et al. (1999) give the density $n=3 \times 10^3 \cc$ and UV
radiation density $G_0=1-3$. The same models estimate a temperature of
about 30-40 K at the surface of the molecular clouds and a thermal
pressure of $\simeq 10^5 K \cc$, which is 30 times higher than the
local solar neighborhood value, but 10 times lower than the pressure
at the center regions of the Milky Way ($10^6 K \cc$: Spergel \& Blitz
1992). This is a plausible, since the LWS beam encompasses gas at
radius $< 2$ kpc from the center of NGC~5866. The PDR models then
overpredict ([CII]+[OI])/FIR $=5 \times 10^{-3}$, when the observed
([CII]+[OI])/FIR $= 2 \times 10^{-3}$. The PDR models assume that equal
amounts of heating of the dust comes from UV and non-UV photons. The
current discrepancy in ([CII]+[OI])/FIR ratio thus implies that four
times as much heating of the dust grains comes from non-UV radiation
as from UV. This is also our favored scenario for the observed low
$\rat$ and $\2rat$.

\subsection{Gas energetics}

In spiral galaxies [CII] is frequently used as a measure of
star-formation activity (Stacey et al. 1991).  In early type galaxies
emission in the fine structure lines [CII] and [OI] do not necessarily
indicate current star-formation (or PDR emission). In spirals, [CII]
and [OI] are the major coolant of warm, neutral gas and most of the
heating is by photoelectrons from dust grains (Watson 1972). The
potential barrier to photoejection from neutral grains is
approximately 6 eV (Bakes \& Tielens 1994), or higher if the grains
are positively charged. This means that only photons shortward of
$\sim 2000 \AA$ contribute to heating of the gas. Old stellar
populations (PNe, HB, PAGB stars) in Es/S0s produce enough ionizing
flux shortward of $912 \AA$ to explain the H$\alpha$ emission seen
(Rose \& Tinsley 1974, Binette et al. 1994) \footnote{If the ionizing
photons come from old stellar populations, one might expect the
H$\alpha$ distribution to be diffusely distributed, but it is usually
seen to be associated with dust lanes (Goudfrooij et al. 1994)}. So
it is possible that they contribute to heating the gas by
photoelectric heating. But what is the magnitude of that contribution?

To answer that question we estimate photoelectric gas heating by UV
photons from old stellar populations.  We use GISSEL (Bruzual and
Charlot 1993) stellar population models of instantaneous burst of ages
10 and 13 Gyr to calculate the UV spectra of early type galaxies. The
mass of each galaxy is normalized to match the B-band flux within the
LWS beam of $70 \arcsec$. An upper limit to the gas heating rate is
calculated by assuming that the work function of the grains is 6 eV
and the yield of photoelectrons heating process is $\gamma=0.1$
(Tielens \& Hollenbach 1985).  This is an upper limit on the heating
and a lower limit on the UV required because it neglects the positive
charging of the grains by photoelectric process, which raises the
potential barrier for ejecting electrons off grains, therefore
requiring even more energetic photons. We neglect here the
contribution of negatively charged grains, which have an ionization
potential as low as 2 eV, because they are estimated to contribute
less than 10\% of the heating (Spaans et al. 1994). The total power
that goes into heating the gas is given by:

$$L_{PE}=\gamma \times f_{abs} \ \int {(h\nu - 6 eV)\times N_{\nu}} \ d\nu$$

In the above equation $ N_{\nu}$ are the number of photons at
frequency $\nu$ and $f_{abs}$ is the fraction of photons absorbed by
the dust. $f_{abs}$ should be less than unity given the low ratio of
ISM to stars. $f_{abs}$ is estimated from the observed values of
$\lb/\lfir$ and using a standard extinction curve with the spectral
energy distribution from stellar population models. We note here in
passing that it was not possible to reproduce the high $\lfir/\lb=0.29$
seen in NGC~1155 with any of the old population models by any
extinction values.  For the other three galaxies A$_V$ varied from
0.05 to 0.25. Table 2 lists the heating luminosity L$_{\rm PE}$ from
old stellar populations, compared to the luminosity of the [CII] and
[OI] lines; for NGC~1052 and NGC~6958 the [OI] line strength is
estimated using the correlation between [OI]/[CII] and $\6f/\f100$
(Malhotra et al., 2000).

\begin{table}
\caption{Luminosities in cooling lines compared to photoelectric heating}
\begin{flushleft}
\begin{tabular}{lccccccl}
\hline\noalign{\smallskip}
Name & L([CII]) \tnm{a} & L([OI]) \tnm{a}& L(CII+OI)& L(H$\alpha$)&$L_{PE}$(10 Gyr)&$L_{PE}$(13 Gyr) \cr
& & &(Energy & in & $10^6) \ \lsun$ & & \cr
\noalign{\smallskip}
\hline\noalign{\smallskip}
NGC~1052 & 0.5&(0.25) \tnm{b} &0.75 & 3.2   & 0.11 & 0.36 \cr
NGC~1155 & 20 & 20    &40   & 8.1   & 0.85 & 2.80 \cr
NGC~5866 & 2 &  0.5   &2.5  & 4.2   & 0.24 & 0.78 \cr
NGC~6958 & 1.6&(0.66) \tnm{b} &2.3  & 0.025 & 0.46 & 1.58 \cr
\noalign{\smallskip}
\hline
\end{tabular}
\tablenotetext{a}{ Fluxes in the [OI](63 $\mum$) and [CII] lines are taken 
from Malhotra et al. 2000, typical uncertainty in a line flux is 30\%}
\tablenotetext{b} { [OI](63 $\mum$) flux is derived from the [CII] flux and
the IRAS colors $F_\nu(60)/F_\nu(100)$ using the correlation seen between 
[OI]/[CII] and $F_\nu(60)/F_\nu(100)$ (Malhotra et al. 2000).}
\end{flushleft}
\end{table}

The UV heating produced by the old stellar populations fails to
account for the [CII] and [OI] emission by factors of 2 or 3 in
NGC~1052, NGC~6958 and NGC~5866, and clearly fails for NGC~1155 by a
factor of about 20. The theoretical estimates of UV production by old
populations are uncertain and vary significantly with age and
metallicity (cf. Magris and Bruzual 1993). Therefore a factor of 2-3
discrepancy in the required UV and theoretical UV production does not
necessarily imply the presence of extra UV sources. However, for
NGC~1155 we can conclude that the strength of [CII] and [OI]
emission imply a higher UV production than can be explained by the old
populations.

 This conclusion is also supported by the optical spectrum of
NGC~1155, which shows narrow emission lines. The ratio of [NII](6583
\AA)/H$\alpha \simeq 0.5$ characteristic of HII regions (Figure
6). NGC~5866 shows a ratio of [NII](6583\AA)/H$\alpha \simeq 1.34$ and
Ho et al. (1997) classify the nucleus as a transition object between a
LINER and an HII nucleus.

\begin{figure}[htb]
\centerline{\epsfxsize=4.7in \epsfbox{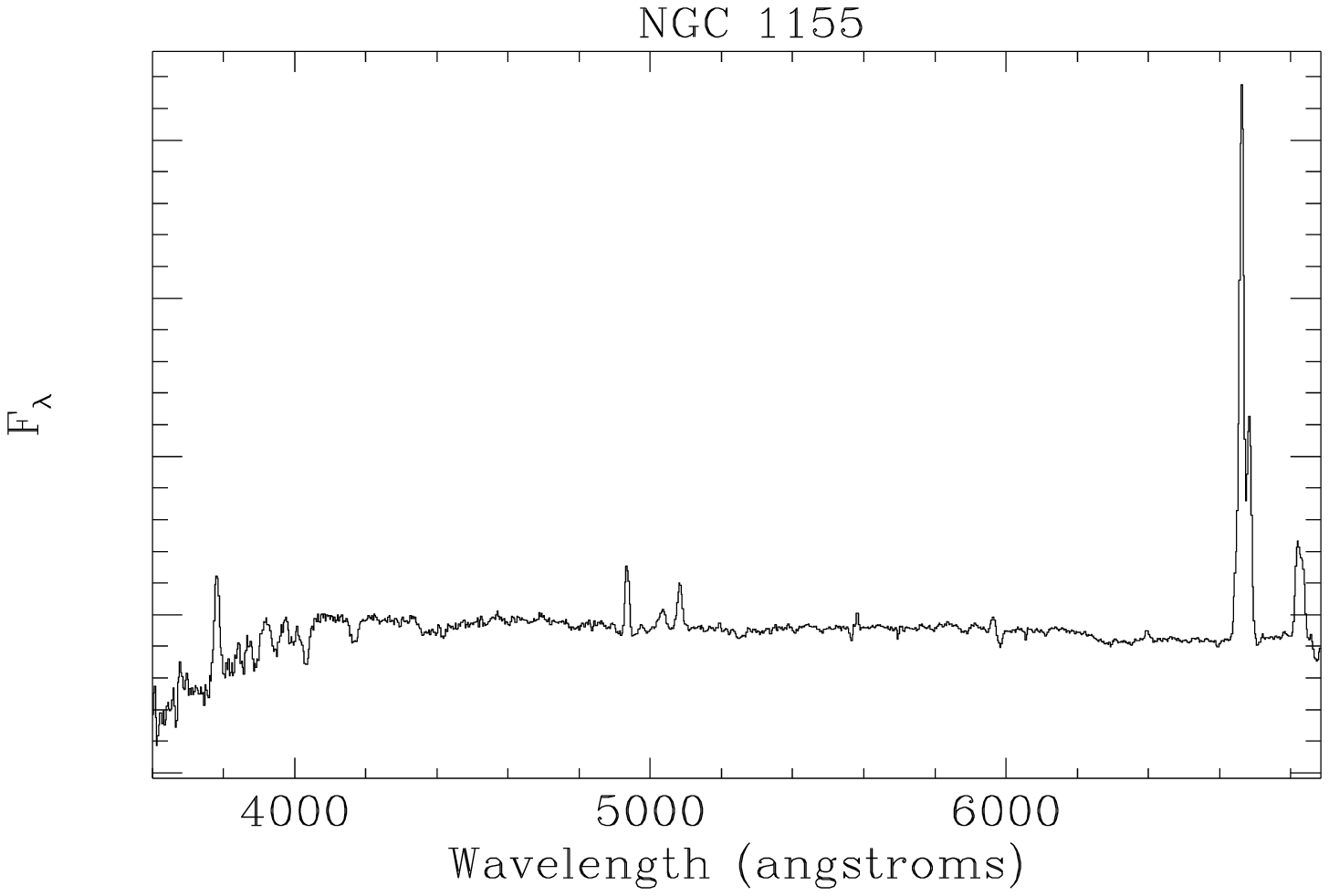} }
\caption{ The optical spectrum on NGC~1155 shows narrow lines and line
ratios characteristic of HII regions; i.e. $H\alpha/[NII]>$1) lending
support to our conclusion from the analysis of FIR lines that there
has been recent star-formation.}
\end{figure}

The observed far-UV-optical colors (1550-V) of Es/S0s vary by a
factor of 10 for active and quiescent galaxies (Brown et al. 1998,
Burstein et al. 1988, hereafter B3FL).  The scatter is reduced by
using a correlation with the magnesium index Mg$_2$. Quiescent
E/S0 galaxies, with or without an active nucleus, show a higher Mg$_2$
index than galaxies with star-formation, except when they have very
red (1550-V) colors. We compare our sample to B3FL figure 1(a), except
we estimate the flux in the 1500\AA\ band from the UV required to
produce the [CII] and [OI] line emission (hereafter referred to as
FUV$_{[CII]}$), since 3 out of 4 galaxies (except NGC~1052) do not
have their far-UV measured. We cannot comment on NGC~1155 whose V or
Mg$_2$ measurements are not available. NGC~6958 and NGC~5866 have low
Mg$_2$ indices, 0.23 and 0.21 respectively (Davies et al. 1987, Huchra
et al. 1996, Fisher et al. 1996). NGC~5866 has normal FUV$_{[CII]}$-V
color so the low Mg$_2$ could be due to recent star-formation, or due
to extinction by the central dust lane (Fisher, Franx \& Illingworth
1996). NGC~6958 has red color FUV$_{[CII]}$-V=4.2, and lies on the
locus of quiescent galaxies with red colors and low Mg$_2$ indices,
indicating little recent star-formation. For NGC~1052 the
FUV$_{[CII]}$-V=1.08 magnitudes, 2 magnitudes bluer than the IUE
observed color (B3FL). In other words, from the [CII] estimate
NGC~1052 is bluer than the sample of quiescent galaxies. The
FUV$_{[CII]}$ is also 6 times higher than the observed UV in this galaxy,
although some uncertainty remains because IUE measured fluxes between
1300 and 3000 \AA\ and FUV$_{[CII]}$ is an estimate of flux between
912 and 2070 \AA\ . The extra UV in NGC~1052 could come from very
obscured source(s), either star-formation or the central AGN.

Other sources of gas heating include X-rays from hot gas and active
nuclei. Only NGC~1052 shows evidence for a Seyfert nucleus and is
detected in X-rays by Einstein and ROSAT. It shows a higher luminosity
in X-rays than the average $L_X-L_B$ relation (Fabbiano et al. 1992)
and the X-ray gas temperature is much higher than expected from the
average relation between stellar velocity dispersion and temperature
of the X-ray gas, indicating that X-ray emission is dominated by AGN
or X-ray binaries rather than diffuse hot gas (Davis and White
1996). The X-ray luminosity of NGC~1052 is $\simeq 3 \times 10^6
\lsun$. About 10\% of X-ray luminosity could be used for gas heating
in atomic medium and about 40\% in molecular medium in the so called
X-ray dissociation regions XDRs (Maloney et al. 1996). In NGC~1052
X-rays could make a significant contribution to gas heating. But in
XDRs one expects a higher line to continuum ratio for the [CII] and
[OI] lines because the X-rays are equally efficient at heating the gas
and the dust, whereas the UV radiation in PDRs heats dust much more
efficiently. Because NGC~1052 shows a lower $\rat$ than the average
for other (mostly spiral) galaxies in the sample, a significant X-ray
heating of the cool gas seems unlikely. NGC~1155 and NGC~6958 were not
detected by Einstein or ROSAT all sky survey. We estimate the X-ray
luminosity from the $L_X-L_B$ relation (Fabbiano et al. 1992) to be
$\simeq 10^5 \lsun$ and $\simeq 0.7 \times 10^5 \lsun$ for NGC~6958
and NGC~1155 respectively. That is simply not sufficient to heat the
gas to produce the [CII] flux seen (Table 2). NGC~5866 has a
luminosity of $L_X (N5866)=0.7\times10^5 \lsun$, again insufficient to
heat the cool gas enough to explain the flux in the far-infrared
cooling lines.

\section{Summary and Conclusions}

From observations of two elliptical galaxies and two S0s with ISO-CAM
and ISO-LWS we can conclude that

(1) The Mid-infrared emission from early type galaxies arises both
from ISM and photospheres of stars. 7/15 $\mum$ ratio varies with
radius indicating that ISM dominates at the centers and in NGC~6958 is
not very extended. In NGC~1155 the dust emission extends to 10 Kpc. In
NGC~1052 the Seyfert nucleus dominates the mid-IR emission. In
NGC~5866, the mid-IR emission at 7 and $15 \mum$ follows the central
dust lane.

(2) [CII] (158 $\mum$) and [OI] (63 $\mum$) lines are detected in
 galaxies where \HI\ and CO detections were difficult, making ISO-LWS
 one of the sensitive probes of small quantities of cool ISM (T $\sim
 100$ K).  [CII] is not mostly from classical HII regions. [CII]
 emission is associated with \HI\ in NGC~1052, and with molecular gas
 in NGC~5866.

(3) The line to continuum ratio $\rat$ measures the efficiency of gas
heating by photoelectric heating.  In three of the four early type
galaxies considered here $\rat$ is lower by a factor of 2-5 than the
typical values in a sample of 60 normal galaxies. A softer radiation
field, i.e. relatively deficient in UV is the most likely
explanation. This is corroborated by a low $\2rat$ ratio observed in
NGC~5866.

(4) The fluxes in the cooling lines, [CII] and [OI], provide fairly
stringent lower limits on the UV flux between 912 and $2100 \AA$. With
all the uncertainties in modelling, UV from evolved stellar
populations is insufficient by a factor of 2-3 for NGC~1052,
NGC~5866 and NGC~6958 and a clear factor of 20 for for
NGC~1155. Heating of cool ISM by X-rays is a plausible scenario only
for NGC~1052. In NGC~6958, NGC~1052 and NGC~5866 it is plausible that
old stars produce enough UV to heat the ISM to explain the cooling by
[CII] line. In NGC~1155, young stars are needed to provide the UV
heating to account for the cooling via the FIR lines.

(5) Comparing the [CII] and [OI] ($63\mum$) line flux measurements of
NGC~1155 with PDR models by Kaufman et al. 1999 we infer gas density $n
\simeq 10^2 \cc$ and UV radiation density $\simeq 10^2 \ G_0$, where
$G_0$ is the UV radiation density in the solar neighborhood. Comparing
the models to the observed [CII], [OI] and CO(1-0) line ratios for
NGC~5866 yields $G_0=1-3$, and $n=3 \times 10^3 \cc$, and the estimate that
non-UV radiation contributes about 80\% of the dust heating in this
galaxy.

\acknowledgements We thank Michael Kaufman, Harry Ferguson, Sue
Madden, John Mulchaey, Jill Knapp and James Rhoads, for helpful
discussions. We also thank the anoymous referee, for many suggestions
that improved the paper. This research has made use of the NASA/IPAC
Extragalactic Database (NED) which is operated by the JPL, California
Institute of Technology, under contract with the NASA. This work was
supported by ISO data analysis funding from NASA, and carried out at
IPAC and the JPL of the California Institute of Technology.  SM's
research is supported by a Hubble Fellowship grant \# HF-01111.01-98A
from the Space Telescope Science Institute, which is operated by the
Association of Universities for Research in Astronomy, Inc., under
NASA contract NAS5-26555. ISO is an ESA project with instruments
funded by ESA Member States (especially the PI countries: France,
Germany, the Netherlands and the United Kingdom), and with the
participation of ISAS and NASA.

\end{document}